\documentclass[11pt,a4paper]{article}
\usepackage[hyperref]{acl2019}
\usepackage{times}
\usepackage{latexsym}
\usepackage{graphicx}  
\graphicspath{{images/}}
\usepackage{pgfplots}
\pgfplotsset{compat=1.12}
\usepackage{amsmath}
\usepackage{dblfloatfix}
\usepackage{here}
\DeclareMathOperator*{\argmin}{argmin} 
\DeclareMathOperator*{\argmax}{argmax} 
\usepackage{tabu}
\usepackage{longtable}
\usepackage{multirow}
\usepackage{algorithm}
\usepackage{algorithmic}
\usepackage{soul}
\newcommand{\revision}[1]{{\color{black}#1}}
\newcommand{\revise}[1]{{\color{black}#1}}

\usepackage{tikz}
\usepackage{xcolor}
\usepgfplotslibrary{groupplots}
\usepackage{colortbl,array,xcolor, booktabs}
\newcommand{\squishlist}{
 \begin{list}{$\bullet$}
  { \setlength{\itemsep}{0pt}
     \setlength{\parsep}{1pt}
     \setlength{\topsep}{1pt}
     \setlength{\partopsep}{0pt}
     \setlength{\leftmargin}{1.5em}
     \setlength{\labelwidth}{1em}
     \setlength{\labelsep}{0.5em} } }
 \newcommand{\squishend}{\end{list}}

\usepackage{color}

\usepackage{url}

\aclfinalcopy 
\def\aclpaperid{1419} 

\title{Low-resource Deep Entity Resolution with Transfer and Active Learning}

\author{Jungo Kasai\textsuperscript{$\heartsuit$}\thanks{\ \ Work done during summer internship at IBM Research -- Almaden.} \quad\quad\quad Kun Qian\textsuperscript{$\clubsuit$}\quad\quad\quad Sairam Gurajada\textsuperscript{$\clubsuit$} \\
{\bf Yunyao Li\textsuperscript{$\clubsuit$}} \quad\quad\quad {\bf Lucian Popa\textsuperscript{$\clubsuit$}}\\
\textsuperscript{$\heartsuit$}Paul G.\ Allen School of Computer Science \& Engineering, University of Washington  \\
\textsuperscript{$\clubsuit$}IBM Research -- Almaden\\
{\tt jkasai@cs.washington.edu}\\
{\tt \{qian.kun,Sairam.Gurajada\}@ibm.com}\\
{\tt \{yunyaoli,lpopa\}@us.ibm.com}}

\date{}

\begin{document}
\maketitle

\begin{abstract}
Entity resolution (ER) is the task of identifying different representations of the same real-world entities across databases. 
It is a key step for knowledge base creation and text mining. 
Recent adaptation of deep learning methods for ER mitigates the need for dataset-specific feature engineering by constructing distributed representations of entity records. While these methods achieve state-of-the-art performance over benchmark data, they require large amounts of labeled data, which are typically unavailable in realistic ER applications. In this paper, we develop a deep learning-based method that targets low-resource settings for ER through a novel combination of transfer learning and active learning. We design an architecture that allows us to learn a transferable model from a high-resource setting to a low-resource one. 
To further adapt to the target dataset, we incorporate active learning that carefully selects a few 
informative examples
to fine-tune the transferred model. Empirical evaluation demonstrates that our method achieves comparable, if not better, performance compared to state-of-the-art learning-based methods while using an order of magnitude fewer labels.
\end{abstract}

\section{Introduction}
Entity Resolution (ER), also known as entity matching, record linkage \cite{Fellegietal1969}, reference reconciliation \cite{Dong:2005:RRC:1066157.1066168}, and merge-purge \cite{Hernandez:1995:MPL:223784.223807}, identifies and links different representations of the same real-world entities.
ER yields a unified and consistent view of data and serves as a crucial step in downstream applications, including knowledge base creation, text mining \cite{zhao-EtAl:2014:Coling1}, and social media analysis \cite{Campbell2016CrossDomainER}.
For instance, seen in Table \ref{example} are citation data records from two databases, DBLP and Google Scholar.
If one intends to build a system that analyzes citation networks of publications, it is essential to recognize publication overlaps across the databases and to integrate the data records \cite{Pasula2002IdentityUA}.

\begin{table*}[t]
\footnotesize
\centering
\begin{tabular}{p{4.7cm}|p{4.15cm}|p{4.5cm}|l}
\hline
\multicolumn{4}{c}{DBLP} \\
  Authors &  Title& Venue&Year\\\hline
  \textcolor{red}{M Carey, D Dewitt, J Naughton, M Asgarian, P Brown, J Gehrke, D Shah}&\textcolor{red}{The Bucky Object-relational Benchmark (Experience Paper)} & \textcolor{red}{SIGMOD Conference}&\textcolor{red}{1997}\\
  \textcolor{blue}{A Netz, S Chaudhuri, J Bernhardt, U Fayyad}&\textcolor{blue}{Integration of Data Mining with Database Technology}&\textcolor{blue}{VLDB}&\textcolor{blue}{2000}\\\hline
  
\multicolumn{4}{c}{Google Scholar} \\
  Authors &  Title& Venue&Year\\\hline
  \textcolor{red}{MJ Carey, DJ Dewitt, JF Naughton, M Asgarian, P  }&\textcolor{red}{The Bucky Object Relational Benchmark}&\textcolor{red}{
  Proceedings of 
  the SIGMOD Conference on Management of Data}&\textcolor{red}{NULL}\\
   \textcolor{brown}{A Netz, S Chaudhuri, J Bernhardt, U Fayyad}&\textcolor{brown}{Integration of Data Mining and Relational Databases}&\textcolor{brown}{Proc.}&\textcolor{brown}{2000}\\\hline
\end{tabular}
\vspace{-1mm}
\caption{Data record examples from DBLP-Scholar (citation genre). The first records from DBLP and Google Scholar (red) refer to the same publication even though the information is not identical. 
The second ones (blue and brown) record different papers with the same authors and year. 
}
\label{example}
\vspace{-2mm}
\end{table*}

\looseness=-1 Recent work demonstrated that deep learning (DL) models with distributed representations of words are viable alternatives to other machine learning algorithms, including support vector machines and decision trees, for performing ER \cite{Ebraheemetal2018,Mudgal2018}.
The DL models provide a universal solution to ER across all kinds of datasets that alleviates the necessity of expensive feature engineering, in which a human designer explicitly defines matching functions for every single ER scenario.
\revise{However, DL is well known to be data hungry; in fact, 
the DL models proposed in \citet{Ebraheemetal2018,Mudgal2018} achieve state-of-the-art performance 
by learning from thousands of labels.\footnote{17k labels were used for the DBLP-Scholar scenario.} 
Unfortunately, realistic ER tasks have limited access to labeled data and would require 
substantial labeling effort upfront, before the actual learning of the ER models. 
Creating a representative training set is especially challenging in ER problems due to the data distribution, 
which is heavily skewed towards negative pairs (i.e.\ non-matches) as opposed to positive pairs (i.e.\ matches).}


\looseness=-1 This problem limits the applicability of DL methods in low-resource ER scenarios. Indeed, we will show in a later section that the performance of DL models degrades significantly as compared to other machine learning algorithms when only a limited amount of labeled data is available. 
To address this issue, we propose a DL-based method that combines transfer learning and active learning. We first develop a transfer learning methodology to leverage a few pre-existing scenarios with abundant labeled data, in order to use them in other  
settings of similar nature but with limited or no labeled data. More concretely, through a carefully crafted neural network architecture, we learn a transferable model from multiple source datasets with cumulatively abundant labeled data.
Then we use active learning to identify informative examples from the target dataset to further adapt the transferred model to the target setting. This novel combination of transfer and active learning in ER settings enables us to learn a comparable or better performing DL model while using significantly fewer target dataset labels in comparison to state-of-the-art \revise{DL and even non-DL} models.
We also note that the two techniques are not dependent on each other. For example, one could 
skip transfer learning if no high-resource dataset is available and directly use active learning. Conversely, one could use transfer learning directly without active learning. We evaluate these cases in the experiments. 
\revise{
\looseness=-1 Specifically, we make the following contributions:
\squishlist
\item We propose a DL architecture for ER that learns attribute agnostic and transferable representations from multiple source datasets using dataset (domain) adaptation.
\item To the best of our knowledge, we are the first to design an active learning algorithm for deep ER models. Our active learning algorithm searches for high-confidence examples and uncertain examples, which provide a guided way to improve the precision and recall of the transferred model to the target dataset. 
\item We perform extensive empirical evaluations over multiple benchmark datasets and demonstrate that our method outperforms state-of-the-art learning-based models while using an order of magnitude fewer labels.
\squishend
}

\section{Background and Related Work}
\subsection{Entity Resolution}
\looseness=-1 \revision{Let $D_1$ and $D_2$ be two collections of entity records. The task of ER is to classify the entity record pair $\langle e_1, e_2\rangle$, $\forall e_1 \in D_1, e_2\in D_2$, into a match or a non-match.
This is accomplished by comparing entity record $e_1$ to $e_2$ on their corresponding attributes. 
In this paper, we assume records in $D_1$ and $D_2$ share the same schema (set of attributes). In cases where 
they have different attributes, one can use schema matching techniques \cite{Rahm2001ASO} to first align the schemas, followed by data exchange techniques \cite{Fagin2009ClioSM}.
Each attribute value is a sequence of words.
Table \ref{example} shows examples of data records from an ER scenario, DBLP-Scholar \cite{Kopcke:2010:EER:1920841.1920904} from the citation genre and clearly depicts our assumption of datasets handled in this paper.
}


\looseness=-1 Since the entire Cartesian product $D_1 \times D_2$ often becomes large and it is infeasible to run a high-recall classifier directly, we typically decompose the problem into two steps: \textit{blocking} and \textit{matching}.
\textit{Blocking} filters out obvious non-matches from the Cartesian product to obtain a candidate set.
\revision{Attribute-level or record-level} tf-idf and jaccard similarity can be used for blocking criteria. For example, in the DBLP-Scholar scenario, one blocking condition could be based on applying equality on ``Year". Hence, two publications in different years will be considered as obvious non-matches and filtered out from the candidate set. 
Then, the subsequent \textit{matching} phase classifies the candidate set into matches and non-matches.

\subsection{Learning-based Entity Resolution}
As described above, after the blocking step, ER reduces to a binary classification task on candidate pairs of data records.
Prior work has proposed learning-based methods that train classifiers on training data, such as support vector machines, naive bayes, and decision trees \cite{Christen:2008:FFA:1385089.1385094,Bilenko:2003:ADD:956750.956759}. 
These learning-based methods first extract features for each record pair from the candidate set across attributes in the schema, and use them to train a binary classifier.
The process of selecting appropriate classification features is often called feature engineering and it involves substantial human effort in each ER scenario. 
Recently, \citet{Ebraheemetal2018} and \citet{Mudgal2018} have proposed deep learning models that use distributed representations of entity record pairs for classification. \revise{These models benefit from distributed representations of words and learn complex features automatically without the need for dataset-specific feature engineering.}



\section{Deep ER Model Architecture}
\begin{figure}
    \centering
    \includegraphics[width=0.46\textwidth]{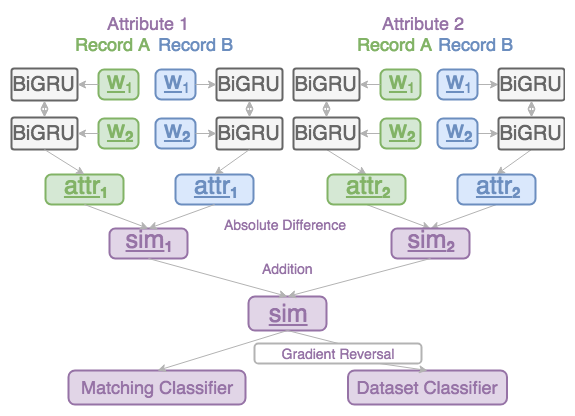}
\vspace{-3mm}
    \caption{Deep ER model architecture with dataset adaptation via gradient reversal. Only two attributes are shown. $W$s indicate word vectors.} 
    \label{deepER}
\end{figure}

%
%
We describe the architecture of our DL model that classifies each record pair in the candidate set into a match or a non-match. As shown in Fig.\ \ref{deepER}, our model encompasses a sequence of steps that computes  attribute representations, attribute similarity and finally the record similarity for each input pair $\langle e_1,e_2\rangle$. A matching classifier uses the record similarity representation to classify the pair. For an extensive list of hyperparameters and training details we chose, see the appendix.




\medskip
\noindent \textbf{Input Representations}.
\revision{For each entity record pair $\langle e_1,e_2\rangle$, we tokenize the attribute values and vectorize the words by external word embeddings to obtain input representations ($W$s in Fig.\ \ref{deepER}).}
We use the 300 dimensional fastText embeddings \cite{Bojanowski2017EnrichingWV}, which capture subword information by producing word vectors via character n-grams. 
This vectorization has the benefit of well representing out-of-vocabulary words \cite{Bojanowski2017EnrichingWV} that frequently appear in ER attributes. For instance, venue names  \textit{SIGMOD} and \textit{ACL} are out of vocabulary in the publicly available GloVe vectors \cite{Pennington2014GloveGV}, but we clearly need to distinguish them.


\medskip
\noindent \textbf{Attribute Representations}.
We build a \revise{universal} bidirectional RNN on the word input representations of each attribute value and obtain attribute vectors ($attr_1$ and $attr_2$ in Fig.\ \ref{deepER}) by concatenating the last hidden units from both directions. Crucially, the universal RNN allows for transfer learning between datasets of different schemas without error-prone schema mapping.
We found that gated recurrent units (GRUs, \citet{Cho2014LearningPR}) yielded the best performance on the dev set as compared to simple recurrent neural networks (SRNNs, \citet{Elman1990FindingSI}) and Long Short-Term Memory networks (LSTMs, \citet{Hochreiter1997LongSM}).
We also found that \revision{using BiGRU with multiple layers} did not help,
and we will use one-layer BiGRUs with 150 hidden units throughout the experiments below.

%
\revision{
\medskip
\noindent \textbf{Attribute Similarity}. The resultant attribute representations are then used to compare attributes of each entity record pair.
In particular, we compute the element-wise absolute difference between the two attribute vectors for each attribute and construct attribute similarity vectors ($sim_1$ and $sim_2$ in Fig.\ \ref{deepER}).
We also considered other comparison mechanisms such as concatenation and element-wise multiplication, but we found that absolute difference performs the best in development, and we will report results from absolute difference.}

\revision{
\medskip
\noindent \textbf{Record Similarity}.
Given the attribute similarity vectors, we now combine those vectors to represent the similarity between the input entity record pair.
Here, we take a simple but effective approach of adding all attribute similarity vectors ($sim$ in Fig.\ \ref{example}).
This way of combining vectors ensures that the final similarity vector is of the same dimensionality regardless of the number of attributes and facilitates transfer of all the subsequent parameters. For instance, the DBLP-Scholar and Cora\footnote{\url{http://www.cs.umass.edu/âŒmccallum/data/cora-refs.tar}} datasets have four and eight attributes respectively, but the networks can share all weights and biases between the two.
We also tried methods such as max pooling and average pooling, but none of them outperformed the simple addition method.}


\medskip
\noindent \textbf{Matching Classification}.
We finally feed the similarity vector for the two records to a two-layer multilayer perceptron (MLP) with highway connections \cite{Srivastava2015TrainingVD} and classify the pair into a match or a non-match (``Matching Classifier" in Fig.\ \ref{deepER}).
The output from the final layer of the MLP is a two dimensional vector and we normalize it by the softmax function to obtain a probability distribution. 
\revision{We 
will discuss 
dataset adaptation for transfer learning 
in 
the next section.}

\medskip
\noindent \textbf{Training Objectives}.
We train the networks to minimize the negative log-likelihood loss.
We use the Adam optimization algorithm \cite{Kingma2015} with batch size 16 and an initial learning rate of 0.001, and after each epoch we evaluate our model on the dev set. Training terminates after 20 epochs, and we choose the model that yields the best F1 score on the dev set and evaluate the model on the test data.

\section{Deep Transfer Active Learning for ER}
We introduce two orthogonal frameworks for our deep ER models in low resource settings: transfer and active learning. 
We also introduce the notion of \textit{likely false positives} and \textit{likely false negatives}, and provide a principled active labeling method in the context of deep ER models, which contributes to stable and high performance.
\subsection{Adversarial Transfer Learning}

\revise{The architecture described above allows for simple transfer learning:} we can train all parameters in the network on source data and use them to classify a target dataset.
However, this method of transfer learning can suffer from dataset-specific properties.
For example, the author attribute in the DBLP-ACM dataset contains first names while that in the DBLP-Scholar dataset only has first initials.
In such situations, it becomes crucial to construct network representations that are invariant with respect
to idiosyncratic properties of datasets.
To this end, we apply the technique of dataset (domain) adaptation developed in image recognition \cite{ganin15}.
In particular,  we build a dataset classifier with the same architecture as the matching classifier (``Dataset Classifier" in Fig.\ \ref{deepER}) that predicts which dataset the input pair comes from.
We replace the training objective by the sum of the negative log-likelihood losses from the two classifiers.
We add a gradient reversal layer between the similarity vector and the dataset classifier so that the parameters in the dataset classifier are trained to predict the dataset while the rest of the network is trained to mislead the dataset classifier, thereby developing dataset-independent internal representations.
Crucially, with dataset adaptation, we feed pairs from the target dataset as well as the source to the network.
For the pairs from the target, we disregard the loss from the matching classifier.
\subsection{Active Learning}
Since labeling a large number of pairs for each ER scenario clearly does not scale, prior work in ER has adopted active learning as a more guided approach to select examples to label \cite{Tejada2001LearningOI,Sarawagi2002InteractiveDU,Arasu2010OnAL,Freitas2010ActiveLG,Isele2013ActiveLO,Qian:2017:ALL:3132847.3132949}.

\revise{Designing an effective active learning algorithm for deep ER models is particularly challenging because finding informative examples is very difficult (especially for positive examples due to the extremely low matching ratio in realistic ER tasks), and we need more than a handful of both negative and positive examples in order to tune a deep ER model with many parameters. 
}

\looseness=-1 \revise{To address this issue, we design an iterative active learning algorithm (Algorithm \ref{active_framework}) that searches for two different types of examples from unlabeled data in each iteration:
(1) uncertain examples including likely false positives and likely false negatives, which will be labeled by human annotators; (2) high-confidence examples including high-confidence positives and high-confidence negatives. We will not label high-confidence examples and use predicted labels as a proxy.
We will show below that those carefully selected examples serve different purposes.}


\looseness=-1 \revision{Uncertain} examples and high-confidence examples are characterized by the entropy of the conditional probability distribution given by the current model.
Let $K$ be the sampling size and the unlabeled dataset consisting of candidate record pairs be $D^U = \{x_i\}_{i=1}^N$.
Denote the probability that record pair $x_i$ is a match according to the current model by $p(x_i)$.
Then, the conditional entropy of the pair $H\left(x_i\right)$ is computed by:
\begin{align*}
-p(x_i) \log p(x_i)-(1-p(x_i))\log(1-p(x_i))
\end{align*}
\revision{Uncertain} examples and high-confidence examples are associated with high and low entropy.

Given this notion of \revision{uncertainty} and high confidence, one can simply select record pairs with top $K$ entropy as \revision{uncertain} examples and those with bottom $K$ entropy as high-confidence examples. Namely, take
\begin{align*}
\argmax_{D \subseteq D^U |D|=K} \sum_{x \in D} H(x),
\argmin_{D \subseteq D^U |D|=K} \sum_{x \in D} H(x)
\end{align*}
as sets of \revision{uncertain} and high-confidence examples respectively.
However, these simple criteria can introduce an unintended bias toward a certain direction, resulting in unstable performance.
For example, \revision{uncertain} examples selected solely on the basis of entropy can sometimes contain substantially more negative examples than positive ones, leading the network to a solution with low recall.
To address this instability problem, we propose a partition sampling mechanism.
We first partition the unlabeled data $D^U$ into two subsets: $\overline{D}^U$ and $\underline{D}^U$, consisting of pairs that the model predicts as matches and non-matches respectively.
Namely,
$\overline{D}^U = \{x\in D^U| p(x) \geq 0.5\},
\underline{D}^U = \{x\in D^U |p(x) <0.5\}$.

Then, we pick top/bottom $k=K/2$ examples from each subset with respect to entropy.
\revision{Uncertain} examples are now:
\begin{align*}
\argmax_{D \subseteq \overline{D}^U |D|=k} \sum_{x \in D} H(x), \argmax_{D \subseteq \underline{D}^U |D|=k} \sum_{x \in D} H(x)
\end{align*}
where the two criteria select \textit{likely false positives} and \textit{likely false negatives} respectively. 
\revise{Likely false positives and likely false negatives are useful for improving the precision and recall of ER models \cite{Qian:2017:ALL:3132847.3132949}. However, the deep ER models do not have explicit features, and thus we use entropy to identify the two types of examples in contrast to the feature-based method used in \citet{Qian:2017:ALL:3132847.3132949}}.
\looseness=-1 High-confidence examples are identified by:
\begin{align*}
\argmin_{D \subseteq \overline{D}^U |D|=k} \sum_{x \in D} H(x),\, \argmin_{D \subseteq \underline{D}^U |D|=k} \sum_{x \in D} H(x)
\end{align*}
where the two criteria correspond to \textit{high-confidence positives} and \textit{high-confidence negatives} respectively.
These sampling criteria equally partition \revision{uncertain} examples and high-confidence examples into different categories.
We will show that the partition mechanism contributes to stable and better performance in a later section.
\begin{algorithm}[htb]
\caption{Deep Transfer Active Learning}
\label{active_framework}
\begin{algorithmic}[1]
\small
\REQUIRE ~~\\
Unlabeled data ${D}^U$, sampling size $K$, batch size $B$, max. iteration number $T$, max. number of epochs $I$.
\ENSURE ~~\\
\looseness=-1 Denote the deep ER parameters and the set of labeled examples by $\mathcal{W}$ and $D^{L}$ respectively.
$\text{Update}(\mathcal{W}, D^L, B)$ denotes a parameter update function that optimizes the negative log-likelihood of the labeled data $D^L$ with batch size $B$. Set $k=K/2$.
\STATE {Initialize $\mathcal{W}$ via transfer learning. Initialize also $D^L=\emptyset$}
\FOR{ { $t \in \left\{1,2,...,T\right\}$}}
\STATE {Select $k$ likely false positives and $k$ likely false negatives from $D^U$ and remove them from $D^U$. Label those examples and add them to $D^L$.
\STATE {Select $k$ high-confidence positives and $k$ high-confidence negatives from $D^U$ and add them with positive and negative labels to $D^L$.}}
\FOR{ {$t \in \left\{1,2,...,I\right\}$}}
\STATE {$\mathcal{W}\gets \text{Update}(\mathcal{W}, D^L, B)$}
\STATE {Run deep ER model on $D^L$ with $\mathcal{W}$ and get the F1 score.}
\IF {the F1 score improves}
\STATE {$\mathcal{W}_{best} \gets \mathcal{W}$}
\ENDIF
\ENDFOR
\STATE {$\mathcal{W} \gets \mathcal{W}_{best}$}
\ENDFOR
\RETURN $\mathcal{W}$
\end{algorithmic}
\end{algorithm}

\revise{
High-confidence examples prevent the network from overfitting to selected uncertain examples \cite{Wang:2017:CAL:3203306.3203314}.
Moreover, they can give the DL model more labeled data without actual manual effort.}
Note that we avoid using any entropy level thresholds to select examples, and instead fix the number of examples.
In contrast, the active learning framework for neural network image recognition in \citet{Wang:2017:CAL:3203306.3203314} uses entropy thresholds.
Such thresholds necessitate fine-tuning for each target dataset: \citet{Wang:2017:CAL:3203306.3203314} use different thresholds for different image recognition datasets.
However, since we do not have sufficient labeled data for the target in low-resource ER problems, the necessity of fine-tuning thresholds would undermine the applicability of the active learning framework.



\section{Experiments}
\subsection{Experimental Setup}
For all datasets, we first conduct blocking to reduce the Cartesian product to a candidate set.
Then, we randomly split the candidate set into training, development, and test data with a ratio of 3:1:1.
For the datasets used in \citet{Mudgal2018} (DBLP-ACM, DBLP-Scholar, Fodors-Zagats, and Amazon-Google), we adopted the same feature-based blocking strategies and random splits to ensure comparability with the state-of-the-art method.
The candidate set of Cora was obtained by randomly sampling 50,000 pairs from the result of the jaccard similarity-based blocking strategy described in \citet{Wang:2011:EMS:2021017.2021020}.
The candidate set of Zomato-Yelp was taken from \citet{magellandata}.\footnote{\looseness=-1 We constructed Zomato-Yelp by merging Restaurants 1 and 2, which are available in \citet{magellandata}. Though the two datasets share the same source, their schemas slightly differ: Restaurants 1 has an address attribute that contains zip code, while Restaurants 2 has a zip code attribute and an address attribute. We put a null value for the zip code attribute in Restaurants 1 and avoid merging errors.}
All dataset statistics are given in Table \ref{stat}.
For evaluation, we compute precision, recall, and F1 score on the test sets.
In the active learning experiments, we hold out the test sets \textit{a priori} and sample solely from the training data to ensure fair comparison with non-active learning methods. The sampling size $K$ for active learning is 20. 
As preprocessing, we tokenize with NLTK \cite{nltk} and lowercase all attribute values.
For every configuration, we run experiments with 5 random initializations and report the average. Our DL models are all implemented using the publicly available deepmatcher library.\footnote{\url{https://github.com/anhaidgroup/deepmatcher}}

\renewcommand{\tabcolsep}{4.30pt}
\begin{table}
\centering
\small
\begin{tabular}{ l |ccccc}
\hline
  dataset & genre & size & matches & attr \\\hline
  DBLP-ACM &citation&12,363 &
2,220&4\\
  DBLP-Scholar&citation&28,707&5,347&4\\
  Cora &citation&50,000&3,969&8\\\hline
  	Fodors-Zagats&restaurant&946&110&6\\
  	Zomato-Yelp&restaurant&894&214&4\\
  	Amazon-Google &software&11,460&1,167&3\\\hline
\end{tabular}
\vspace{-1mm}
\caption{Post-blocking statistics of the ER datasets we used. (attr denotes the number of attributes.)}
\label{stat}
\vspace{-3mm}
\end{table}

\subsection{Baselines}
We establish baselines using a state-of-the-art learning-based ER package, Magellan \cite{Konda2016MagellanTB}.
We experimented with the following 6 learning algorithms: Decision Tree, SVM, Random Forest, Naive Bayes, Logistic Regression, and Linear Regression.
We use the same feature set as in \citet{Mudgal2018}.
See the appendix for extensive lists of features chosen.
\subsection{Results and Discussions}
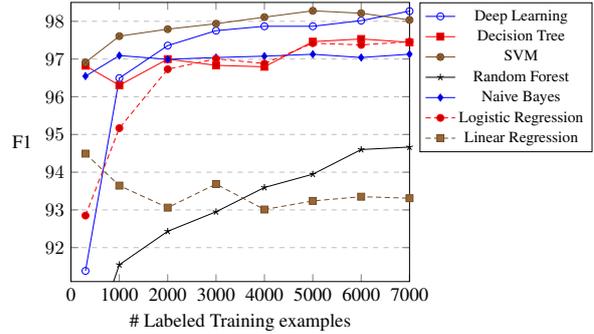
\begin{figure}
\centering
\begin{tikzpicture}[scale=0.65]
\begin{axis}[
    title={},
    xlabel={\# Labeled Training examples},
    ylabel={F1},
    xmin=0, xmax=140,
    ymin=91.1, ymax=98.5,
    xtick={0, 20, 40, 60, 80,100,120, 140},
    xticklabels={0, 1000, 2000, 3000, 4000, 5000, 6000, 7000},
    ytick={92, 93, 94, 95, 96, 97, 98},
    legend pos=outer north east,
    legend style={font=\footnotesize},
    ymajorgrids=true,
    grid style=dashed,
    ylabel style={rotate=-90},
]
 
\addplot
    [
    color=blue,
    mark=o,
    ]
    coordinates {(6,91.382)(20,96.494)(40,97.356)(60,97.752)(80,97.868)(100,97.868)(120,98.018)(140,98.27)};
    \addlegendentry{Deep Learning}
\addplot
    coordinates {(6,96.822)(20,96.304)(40,96.998)(60,96.832)(80,96.794)(100,97.46)(120,97.532)(140,97.442)};
    \addlegendentry{Decision Tree}
\addplot
    coordinates {(6,96.912)(20,97.606)(40,97.792)(60,97.936)(80,98.11)(100,98.278)(120,98.214)(140,98.036)};
    \addlegendentry{SVM}
\addplot
    coordinates {(6,88.944)(20,91.546)(40,92.432)(60,92.95)(80,93.594)(100,93.946)(120,94.6)(140,94.666)};
    \addlegendentry{Random Forest}
\addplot
    coordinates {(6,96.546)(20,97.092)(40,96.992)(60,97.04)(80,97.076)(100,97.126)(120,97.04)(140,97.128)};
    \addlegendentry{Naive Bayes}
\addplot
    coordinates {(6,92.85)(20,95.168)(40,96.73)(60,97.006)(80,96.88)(100,97.42)(120,97.374)(140,97.46)};
    \addlegendentry{Logistic Regression}
\addplot
    coordinates {(6,94.492)(20,93.644)(40,93.06)(60,93.686)(80,93.01)(100,93.238)(120,93.35)(140,93.31)};
    \addlegendentry{Linear Regression}
\end{axis}
\end{tikzpicture}
\vspace{-0.8cm}
\caption{Performance vs. data size (DBLP-ACM).}
\label{data_size}
\vspace{-0.6cm}
\end{figure}

\medskip
\noindent\textbf{Model Performance and Data Size.}
Seen in Fig.\ \ref{data_size} is F1 performance of different models with varying data size on DBLP-ACM.
The DL model improves dramatically as the data size increases and achieves the best performance among the 7 models when 7000 training examples are available.
In contrast, the other models suffer much less from data scarcity with an exception of Random Forest.
We observed similar patterns in DBLP-Scholar and Cora.
These results confirm our hypothesis that deep ER models are data-hungry and require a lot of labeled data to perform well.

\medskip
\looseness=-1 \noindent \textbf{Transfer Learning}.
Table \ref{transfer} shows results from our transfer learning framework when used in isolation (i.e., without active learning, which we will discuss shortly).
Our dataset adaptation method substantially ameliorates performance when the target is DBLP-Scholar (from 41.03 to 53.84 F1 points) or Cora (from 38.3 to 43.13 F1 points) and achieves the same level of performance on DBLP-ACM.
Transfer learning with our dataset adaptation technique achieves a certain level of performance without any target labels, but we still observe high variance in performance (e.g.\ 6.21 standard deviation in DBLP-Scholar) and a huge discrepancy between transfer learning and training directly on the target dataset.
To build a reliable and stable ER model, a certain amount of target labels may be necessary, which leads us to apply our active learning framework.
\begin{table*}[ht]
\small
\centering
\begin{tabular}{ l| ccc|ccc|ccc}
\hline
\multicolumn{1}{r|}{Target} & \multicolumn{3}{c|}{DBLP-ACM} & \multicolumn{3}{c|}{DBLP-Scholar} &\multicolumn{3}{c}{Cora}\\
  Method &Prec&Recall&F1&Prec&Recall&F1&Prec&Recall&F1 \\\hline
  Train on Source &$86.98$&$98.38$&$92.32_{\pm 1.15}$&$73.41$&$43.20$&$41.03_{\pm 6.33}$&$92.54$&$24.22$&$38.30_{\pm 3.77}$\\
  +Adaptation &$88.71$&$96.21$&$92.31_{\pm 1.36}$&$88.06$&$39.03$&$\mathbf{53.84_{\pm 6.21}}$&$40.64$&$52.16$&$\mathbf{43.13_{\pm 3.62}}$\\\hline
   
  Train on Target&$98.30$&$98.60$&$98.45_{\pm 0.22}$&$92.72$&$93.08$&$92.94_{\pm 0.47}$&$98.01$&$99.37$&$98.68_{\pm 0.26}$\\
  \citet{Mudgal2018} & -- & -- & 98.4 & --&--&93.3  & -- & -- & --
\end{tabular}
\vspace{-3mm}
\caption{Transfer learning results (citation genre). We report standard deviations of the F1 scores. For each target dataset, the source is given by  
the other two datasets (e.g., the source for DBLP-ACM is DBLP-Scholar and Cora.)}
\label{transfer}
\vspace{2mm}
\end{table*}

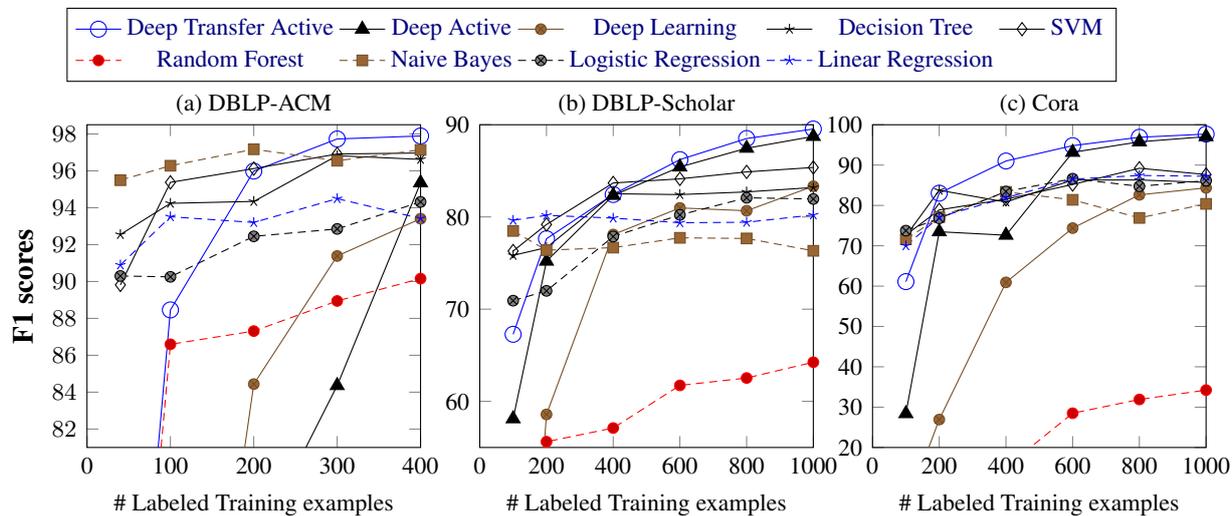
\begin{figure*}[ht]
\hspace{-2.7mm}
\begin{tikzpicture}
\centering
\pgfplotsset{/pgf/number format/fixed, every tick label/.append style={font=\footnotesize}}
  \begin{groupplot}
    [
        anchor=north west,
        group style=
        {
            group size=3 by 1,
            horizontal sep=4.5ex,
            x descriptions at=edge bottom,
            y descriptions at=edge left,
        },
        height=\textwidth/2.73,
        width=\textwidth/2.68
    ]
    \nextgroupplot
        [
            title={\small (a) DBLP-ACM},
            xmin=0, xmax=80,
            ymin=81, ymax=98.5,
            xtick={0, 20, 40, 60, 80},
            xticklabels={0, 100, 200, 300, 400},
            ytick={82, 84, 86, 88, 90, 92, 94, 96, 98},
			legend to name=grouplegend,
			legend style={font=\small},
			legend columns=5,
			y label style={at={(axis description cs:-0.13,.5)}},
			title style={at={(axis description cs:0.5,0.95)}},
			ylabel=\textbf{F1 scores},
			xlabel={\small \# Labeled Training examples},
		]
    \addplot[color=blue,mark=o,mark size=3] 
    coordinates {(8,57.922)(20,88.458)(40,95.986)(60,97.728)(80,97.89)};
    \addlegendentry{Deep Transfer Active}
    \addplot[mark=triangle*, mark size=3]
    coordinates {(8,29.644)(20,49.496)(40,75.412)(60,84.356)(80,95.352)};
    \addlegendentry{Deep Active}
    \addplot
    coordinates {(8,0)(20,52.866)(40,84.436)(60,91.382)(80,93.404)};
    \addlegendentry{Deep Learning}
    \addplot
    coordinates {(8,92.546)(20,94.238)(40,94.342)(60,96.822)(80,96.62)};
    \addlegendentry{Decision Tree}
    \addplot[mark=diamond, mark size=2.5]
    coordinates {(8,89.816)(20,95.376)(40,96.122)(60,96.912)(80,96.966)};
    \addlegendentry{SVM}
    \addplot
    coordinates {(8,56.676)(20,86.594)(40,87.31)(60,88.944)(80,90.156)};
    \addlegendentry{Random Forest}
    \addplot
    coordinates {(8,95.494)(20,96.276)(40,97.17)(60,96.546)(80,97.138)};
    \addlegendentry{Naive Bayes}
    \addplot
    coordinates {(8,90.298)(20,90.252)(40,92.448)(60,92.85)(80,94.316)};
    \addlegendentry{Logistic Regression}
    \addplot
    coordinates {(8,90.892)(20,93.51)(40,93.2)(60,94.492)(80,93.416)};
    \addlegendentry{Linear Regression}

    \nextgroupplot
        [
        title={\small (b) DBLP-Scholar},
        title style={at={(axis description cs:0.5,0.95)}},
        xmin=0, xmax=100,
        ymin=55, ymax=90,
        xtick={0, 20, 40, 60, 80,100},
        xticklabels={0, 200, 400, 600, 800, 1000},
        ytick={60, 70, 80, 90},
        yticklabels={60, 70, 80, 90},
        xlabel={\small \# Labeled Training examples},
        ]
        
    \addplot[color=blue,mark=o,mark size=3] 
    coordinates {(10,67.274)(20,77.636)(40,82.436)(60,86.218)(80,88.504)(100,89.536)};
    \addplot[mark=triangle*, mark size=3]
    coordinates {(10,58.112)(20,75.205)(40,82.338)(60,85.418)(80,87.45)(100,88.732)};
    \addplot
    coordinates {(10,8.866)(20,58.586)(40,78.092)(60,80.98)(80,80.672)(100,83.33)};
    \addplot
    coordinates {(10,75.836)(20,76.676)(40,82.516)(60,82.426)(80,82.714)(100,83.194)};
    \addplot[mark=diamond, mark size=2.5]
    coordinates {(10,76.304)(20,79.18)(40,83.686)(60,84.132)(80,84.886)(100,85.364)};
    \addplot
    coordinates {(10,45.076)(20,55.624)(40,57.11)(60,61.734)(80,62.534)(100,64.246)};
    \addplot
    coordinates {(10,78.482)(20,76.398)(40,76.674)(60,77.736)(80,77.668)(100,76.328)};
    \addplot
    coordinates {(10,70.922)(20,71.984)(40,77.878)(60,80.22)(80,82.078)(100,81.948)};
    \addplot
    coordinates {(10,79.634)(20,80.176)(40,79.874)(60,79.372)(80,79.426)(100,80.214)};

  \nextgroupplot
        [
        title={\small (c) Cora},
        title style={at={(axis description cs:0.5,0.95)}},
        xmin=0, xmax=100,
        ymin=20, ymax=100,
        xtick={0,20, 40, 60, 80,100},
        xticklabels={0, 200, 400, 600, 800, 1000},
        ytick={20, 30, 40, 50, 60, 70, 80, 90, 100},
        yticklabels={20, 30, 40, 50, 60, 70, 80, 90, 100},
        xlabel={\small \# Labeled Training examples},
        ]
    \addplot[color=blue,mark=o,mark size=3] 
    coordinates {(10,61.114)(20,83.036)(40,91.012)(60,94.816)(80,96.896)(100,97.682)};
    \addplot[mark=triangle*, mark size=3]
    coordinates {(10,28.386)(20,73.478)(40,72.614)(60,93.162)(80,95.762)(100,97.054)};
    \addplot
    coordinates {(10,4.936)(20,26.924)(40,60.928)(60,74.38)(80,82.622)(100,84.352)};
    \addplot
    coordinates {(10,72.426)(20,83.758)(40,80.842)(60,86.3)(80,86.308)(100,85.706)};
    \addplot[mark=diamond, mark size=2.5]
    coordinates {(10,72.684)(20,78.892)(40,81.542)(60,85.194)(80,89.212)(100,87.66)};
    \addplot
    coordinates {(10,5.36)(20,11.974)(40,15.148)(60,28.492)(80,31.904)(100,34.222)};
    \addplot
    coordinates {(10,71.588)(20,77.116)(40,83.358)(60,81.39)(80,76.9)(100,80.386)};
    \addplot
    coordinates {(10,73.746)(20,76.802)(40,83.448)(60,86.622)(80,84.742)(100,86.142)};
    \addplot
    coordinates {(10,70)(20,77.126)(40,81.914)(60,86.512)(80,87.428)(100,87.24)};
  \end{groupplot}
  \node at (group c1r1.north east) [inner sep=0pt,anchor=north, yshift=9ex, 
	  	 xshift=13ex] {\ref{grouplegend}};
\end{tikzpicture}
\vspace{-8mm}
\caption{Low-resource performances on different datasets.}
\label{active_plot}
\end{figure*}

\medskip
\looseness=-1 \noindent \textbf{Active Learning}.
Fig.\ \ref{active_plot} shows results from our active learning as well as the 7 algorithms trained on labeled examples of corresponding size that are randomly sampled.\footnote{We average the results over 
5 random samplings.}
Deep transfer active learning (DTAL) initializes the network parameters by transfer learning whereas deep active learning (DAL) starts with a random initialization.
We can observe that DTAL models remedy the data scarcity problem as compared to DL models with random sampling in all three datasets. DAL can achieve competitive performance to DTAL at the expense of faster convergence.

Seen in Table \ref{high_low} is performance comparison of different algorithms in low-resource and high-resource settings. (We only show the SVM results since SVM performed best in each configuration among the 6 non-DL algorithms.)
First, deep transfer active learning (DTAL) achieves the best performance in the low-resource setting of each dataset.
In particular, DTAL outperforms the others to the greatest degree in Cora (97.68 F1 points) probably because Cora is the most complex dataset with 8 attributes in the schema. 
Non-DL algorithms require many interaction features, which lead to data sparsity.
Deep active learning (DAL) also outperforms SVM and yields comparable performance to DTAL. However, the standard deviations in performance of DAL are substantially higher than those of DTAL (e.g.\ 4.15 vs. 0.33 in DBLP-ACM), suggesting that transfer learning provides useful initializations for active learning to achieve stable performance. 

One can argue that DTAL performs best in the low-resource scenario, but the other algorithms can also boost their low-resource performance by active learning.
While there are many approaches to active learning on feature-based (non-DL) ER (e.g.\ \citet{Bellare2012ActiveSF,Qian:2017:ALL:3132847.3132949}) that yield strong performance under certain condition, it requires further research to quantify how these methods perform with varying datasets, genres, and blocking functions.
It should be noted, however, that in DBLP-Scholar and Cora, DTAL in the low-resource setting even significantly outperforms SVM (and the other 5 algorithms) in the high-resource scenario.
These results imply that DTAL would significantly outperform SVM with active learning in the low-resource setting
since the performance with the full training data with labels serves as an upper bound.
Moreover, we can observe that DTAL with a limited amount of data (less than 6\% of training data in all datasets), performs comparably to DL models with full training data.
Therefore, we have demonstrated that a deep ER system with our transfer and active learning frameworks can provide a stable and reliable solution to entity resolution with low annotation effort.

\begin{table}[t]
\footnotesize
\centering
\begin{tabular}{ lcccc}
\bottomrule
  Dataset & Method &  Train Size & F1\\
  [-2pt]
  \bottomrule
   \rowcolor[gray]{.90}&DTAL&400&$\mathbf{97.89_{\pm 0.33}}$\\
  \rowcolor[gray]{.90}&DAL&400&$95.35_{\pm 4.15}$\\
  \rowcolor[gray]{.90}&DL&400&$93.40_{\pm 2.61}$\\
  \rowcolor[gray]{.90}&SVM&400&$96.97_{\pm 0.69}$\\
  &DL&7,417&$98.45_{\pm 0.22}$\\
  \multirow{-6}{*}{DBLP-ACM}&SVM&7,417&$98.35_{\pm 0.14}$\\
  [-2pt]
  \bottomrule
  \rowcolor[gray]{.90}
  &DTAL &1000&$\mathbf{89.54_{\pm 0.39}}$\\
  \rowcolor[gray]{.90}
  &DAL&1000&$88.76_{\pm 0.76}$\\
  \rowcolor[gray]{.90}
  &DL&1000&$83.33_{\pm 1.26}$\\
  \rowcolor[gray]{.90}
  &SVM&1000&$85.36_{\pm 0.32}$\\
  &DL&17,223&$92.94_{\pm 0.47}$\\
  \multirow{-6}{*}{DBLP-Scholar}&SVM&17,223&$88.56_{\pm 0.46}$\\
  [-2pt]
  \bottomrule
  \rowcolor[gray]{.90}
  &DTAL &1000&$\mathbf{97.68_{\pm 0.39}}$\\
  \rowcolor[gray]{.90}
  &DAL&1000&$97.05_{\pm 0.64}$\\
  \rowcolor[gray]{.90}
  &DL&1000&$84.35_{\pm 4.25}$\\
  \rowcolor[gray]{.90}
  &SVM&1000&$87.66_{\pm 3.15}$\\
  &DL&30,000&$98.68_{\pm 0.26}$\\
  \multirow{-6}{*}{Cora}&SVM&30,000&$95.39_{\pm 0.31}$\\
  [-2pt]
  \bottomrule
\end{tabular}
\looseness=-1 \caption{Low-resource (shaded) and high-resource (full training data) performance comparison. DTAL, DAL, and DL denote deep transfer active learning, deep active learning, and deep learning (random sampling).} 
\label{high_low}
\end{table}

\begin{figure}
\vspace{-2mm}
\centering
\begin{tikzpicture}[scale=0.60]
\begin{axis}[
    title={},
    xlabel={\# Labeled Training examples},
    ylabel={F1},
    xmin=0, xmax=140,
    ymin=0, ymax=50,
    xtick={0, 20, 40, 60, 80,100,120,140},
    xticklabels={0, 200, 400, 600, 800, 1000, 1200, 1400},
    ytick={0, 10, 20, 30, 40, 50},
    legend pos=outer north east,
    legend style={font=\footnotesize},
    ymajorgrids=true,
    grid style=dashed,
    ylabel style={rotate=-90},
]
 
\addplot
    [
    color=black,
    mark=o,
    ]
    coordinates {(4,14.24)(20,17.53)(40,20.35)(60,32.57)(80,38.67)(100,44.64)(120,47.3)(140,49.5)};
    \addlegendentry{Deep Active}
\addplot
    coordinates {(4,0)(20,0)(40,11.038)(60,21.068)(80,21.29)(100,25.77)(120,29.32)(140,34.0)};
    \addlegendentry{Deep Learning}
\addplot
    coordinates {(4,25.414)(20,31.528)(40,35.098)(60,33.696)(80,37.716)(100,35.944)(100,36.1)(120,36.3)(140,36.2)};
    \addlegendentry{Decision Tree}
\addplot
    coordinates {(4,15.016)(20,26.838)(40,27.06)(60,29.2)(80,31.444)(100,32.834)(120,34.2)(140,38.2)};
    \addlegendentry{SVM}
\addplot
    coordinates {(4,8.934)(20,18.028)(40,20.272)(60,20.952)(80,26.014)(100,24.3)(120,25.01)(140,25.5)};
    \addlegendentry{Random Forest}
\addplot
    coordinates {(4,20.428)(20,36.176)(40,36.992)(60,38.39)(80,38.998)(100,36.056)(120,36.1)(140,36.3)};
    \addlegendentry{Naive Bayes}
\addplot
    coordinates {(4,29.208)(20,45.31)(40,44.898)(60,45.03)(80,45.908)(100,45.592)(120,45.5)(140,45.3)};
    \addlegendentry{Logistic Regression}
\addplot
    coordinates {(4,21.876)(20,26.164)(40,26.234)(60,27.33)(80,21.124)(100,26.214)(120,28.5)(140,29.0)};
    \addlegendentry{Linear Regression}
\end{axis}
\end{tikzpicture}
\vspace{-4mm}
\caption{Low-resource performance (software genre).}
\label{active_software}
\end{figure}
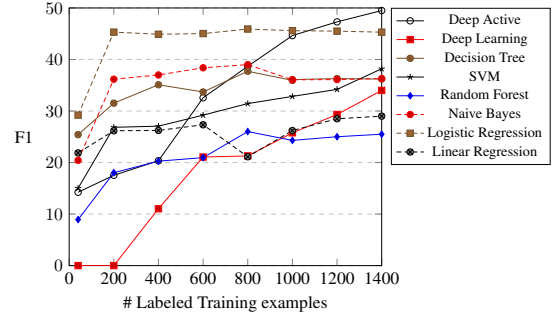

\medskip
\looseness=-1 
\noindent\textbf{Other Genre Results}.
We present results from the \textit{restaurant} and \textit{software} genres.\footnote{We intend to apply our approaches to more genres, but unfortunately we lack large publicly available ER datasets in other genres than citation. Applications to non-English languages are also of interest. We leave this for future.}
Shown in Table \ref{restaurant} are results of transfer and active learning from Zomato-Yelp to Fodors-Zagats.
Similarly to our extensive experiments in the citation genre, the dataset adaptation technique facilitates transfer learning significantly, and only 100 active learning labels are needed to achieve the same performance as the model trained with all target labels (894 labels). 
Fig.\ \ref{active_software} shows low-resource performance in the software genre.
The relative performance among the 6 non-DL approaches differs to a great degree as the best non-DL model is now logistic regression, but deep active learning outperforms the rest with 1200 labeled examples (10.4\% of training data).
These results illustrate that our low-resource frameworks are effective in other genres as well.
\begin{table}
\small
\centering
\begin{tabular}{ l| ccc}
\hline
  Method &Prec&Recall&F1 \\\hline
  Train on Src &$100.00$&$6.37$&$11.76_{\pm 6.84}$\\
  +Adaptation &$95.33$&$57.27$&$70.13_{\pm 19.89}$\\
  +100 active labels &$100.00$&$100.00$&$100.00_{\pm 0.00}$\\\hline
  Train on Tgt &$100.00$&$100.00$&$100.00_{\pm 0.00}$\\
  \citet{Mudgal2018} & -- & -- &100
\end{tabular}
\caption{Transfer and active learning results in the restaurant genre.
The target and source datasets are Fodors-Zagats and Zomato-Yelp respectively.}
\label{restaurant}
\end{table}

\medskip
\looseness=-1 \noindent \textbf{Active Learning Sampling Strategies}.
As discussed in a previous section, we adopted high-confidence sampling and a partition mechanism for our active learning.
Here we analyze the effect of the two methods.
Table \ref{analysis} shows deep transfer active learning performance in DBLP-ACM with varying sampling strategies.
We can observe that high-confidence sampling and the partition mechanism contribute to high and stable performance as well as good precision-recall balance.
Notice that there is a huge jump in recall by adding partition while precision stays the same (row 4 to row 3). This is due to the fact that the partition mechanism succeeds in finding more false negatives.
The breakdown of labeled examples (Table \ref{breakdown}) shows that is indeed the case. 
It is noteworthy that  the partition mechanism lowers the ratio of misclassified examples (FP+FN) in the labeled sample set because partitioning encourages us to choose likely false negatives more aggressively, yet false negatives tend to be more challenging to find in entity resolution due to the skewness toward the negative \cite{Qian:2017:ALL:3132847.3132949}.
We observed similar patterns in DBLP-Scholar and Cora.
\begin{table}[t]
\small
\centering
\begin{tabular}{ l| ccc}
\hline
   Sampling Method&  Prec &Recall &F1\\\hline
  High-Confidence&$93.32$&$97.21$&$95.19_{\pm 2.21}$\\
  Partition&$96.14$&$97.12$&$96.61_{\pm 0.57}$\\
  High-Conf.+Part.&$97.63$&$97.84$&$\mathbf{97.73_{\pm 0.43}}$\\
  Top K Entropy&$96.16$&$89.64$&$92.07_{\pm 9.73}$\\
\end{tabular}
\caption{Low-resource performance (300 labeled examples) of different sampling strategies (DBLP-ACM).}
\label{analysis}
\end{table}
\begin{table}
\vspace{-1mm}
\small
\centering
\begin{tabular}{ c|cccc}
\hline
   Method&FP&  TP &FN&TN\\\hline
   Part&$79.6_{5.9}$&$70.4_{5.9}$&$59.2_{5.6}$&$90.8_{5.6}$\\
  W/o Part &$101.6_{7.7}$&$57.4_{15.9}$&$41.6_{4.4}$&$99.4_{ 22.5}$
\end{tabular}
\caption{Breakdown of 300 labeled samples (uncertain samples) from deep transfer active learning in DBLP-ACM. Part, FP, TP, FN, and TN denote the partition mechanism, false positives, true positives, false negatives, and true negatives respectively.}
\label{breakdown}
\vspace{-2mm}
\end{table}

\section{Further Related Work}
Transfer learning has proven successful in fields such as computer vision and natural language processing, where networks for a target task is pretrained on a source task with plenty of training data (e.g.\ image classification \cite{pmlr-v32-donahue14} and language modeling \cite{Peters2018DeepCW}).
In this work, we developed a transfer learning framework for a deep ER model.
Concurrent work \cite{Thirumuruganathan2018ReuseAA} to ours has also proposed transfer learning on top of the features from distributed representations, but they focused on classical machine learning classifiers (e.g., logistic regression, SVMs, decision trees, random forests) and they did not consider active learning.
Their distributed representations are computed in a ``bag-of-words" fashion, which can make applications to textual attributes more challenging \cite{Mudgal2018}.
Moreover, their method breaks attribute boundaries for tuple representations in contrast to our approach that computes a similarity vector for each attribute in an attribute-agnostic manner.
In a complex ER scenario, each entity record is represented by a large number of attributes, and comparing tuples as a single string can be infeasible.
Other prior work also proposed a transfer learning framework for linear model-based learners in ER \cite{Negahban2012ScalingME}.

\section{Conclusion}
We presented transfer learning and active learning frameworks for entity resolution with deep learning and demonstrated that our models can achieve competitive, if not better, performance as compared to state-of-the-art learning-based methods while only using an order of magnitude less labeled data.
Although our transfer learning alone did not suffice to construct a reliable and stable entity resolution system, it contributed to faster convergence and stable performance when used together with active learning.
These results serve as further support for the claim that deep learning can provide a unified data integration method for downstream NLP tasks.
Our frameworks of transfer and active learning for deep learning models are potentially applicable to low-resource settings beyond entity resolution. 

\section*{Acknowledgments}
We thank Sidharth Mudgal for assistance with the DeepMatcher/Magellan libraries and replicating experiments.
We also thank Vamsi Meduri, Phoebe Mulcaire, and the anonymous reviewers for their helpful feedback. JK was supported by travel grants from the Masason Foundation fellowship.

\bibliography{acl2019}
\bibliographystyle{acl_natbib}
\appendix
\section{Appendices}
\subsection{Deep ER Hyperparameters}
Seen in Table \ref{er-hyp} is a list of hyperparameters for our deep entity resolution models. We use the same hyperparameters regardless of scenario and dataset. 
We initialize the 300 dimensional word embeddings by the character-based pretrained fastText vectors publicly available.\footnote{\url{https://github.com/facebookresearch/fastText}}

\begin{table}[h]
\small
\centering
\begin{tabular}{ |l c|}
\hline
\multicolumn{2}{|c|}{Input Representations}\\
Word embedding size & 300\\
Input dropout rate& 0.0\\
\multicolumn{2}{|c|}{Word-level BiGRU}\\
GRU size & 150\\
\# GRU layers & 1\\
Final ouput & concat\\
\multicolumn{2}{|c|}{Similarity Representations}\\
Attr. sim. & absolute diff. \\
Record sim. & sum\\
\multicolumn{2}{|c|}{Matching Classification}\\
\# MLP layers& 2\\
\# MLP size& 300\\
\# MLP activation& relu\\
Highway Connection& Yes\\
\multicolumn{2}{|c|}{Domain Classification (Adversarial)}\\
\# MLP layers& 2\\
\# MLP size& 300\\
\# MLP activation& relu\\
Highway Connection& Yes\\
\hline
\multicolumn{2}{|c|}{Training}\\
Objective & cross-entropy\\
Batch size & 16\\
\# Epochs & 20\\
Adam \cite{Kingma2015} lrate& 0.001\\
Adam $\beta_1$& 0.9\\
Adam $\beta_2$& 0.999\\
\hline
\end{tabular}
\caption{Deep ER hyperparameters.}
\label{er-hyp}
\end{table}
\subsection{Non-DL Learning Algorithms}
Magellan \cite{Konda2016MagellanTB} is an open-source package that provides state-of-the-art learning-based algorithms for ER.\footnote{\url{https://sites.google.com/site/anhaidgroup/projects/magellan}}
We use the package to run the following 6 learning algorithms for baselines: Decision Tree, SVM, Random Forest, Naive Bayes, Logistic Regression, and Linear Regression.
For each attribute in the schema, we apply the following similarity functions: q-gram jaccard, cosine distance, Levenshtein disntance, Levenshtein similairty, Monge-Elkan measure, and exact matching.

\end{document}